\documentclass[%
prd,a4paper,aps,twocolumn,nofootinbib,nobibnotes,preprintnumbers,superscriptaddress,floatfix
]{revtex4-2}

\newcommand\tcut{t_\mathrm{cut}}
\newcommand\scale{M}
\newcommand\klat{\hat k}
\newcommand\kco{\vphantom{\hat k}k}
\newcommand\mlat{\hat m}
\newcommand\llat{\hat \lambda}
\newcommand\mulat{\hat \mu}
\newcommand{\lint}[1]{\int\limits_{#1} }
\newcommand{\cont}{\mathrm{c}}
\newcommand{\Lwall}{L_{\mathrm w}}

\usepackage{graphicx}
\usepackage{amsmath, amssymb}
\usepackage{bm}
\usepackage{hyperref}
\usepackage{color}
\usepackage{siunitx}
\usepackage{slashed}
\usepackage{csquotes}
\usepackage{bm}
\usepackage{mathtools}
\usepackage[compat=1.1.0]{tikz-feynman}
\usetikzlibrary{backgrounds, shapes.multipart, math}
\usepackage{tabularx}
\usepackage{booktabs}

\begin{document}

\preprint{
DESY-26-076, HIP-2026-10/TH
}

\title{Seeded bubble nucleation on the lattice}

\author{Simone Blasi}
\email{simone.blasi@desy.de}
\affiliation{Deutsches Elektronen-Synchrotron DESY, Notkestr.~85, 22607 Hamburg, Germany}

\author{Andreas Ekstedt}
\affiliation{Department of Physics and Astronomy, Uppsala University,
P.O. Box 256, SE-751 05 Uppsala, Sweden}

\author{Jaakko Hällfors}
\email{jaakko.hallfors@helsinki.fi}
\affiliation{Department of Physics and Helsinki Institute of Physics, P.O.\,Box 64, FI-00014 University of Helsinki, Finland}

\author{Kari Rummukainen}
\email{kari.rummukainen@helsinki.fi}
\affiliation{Department of Physics and Helsinki Institute of Physics, P.O.\,Box 64, FI-00014 University of Helsinki, Finland}

\date{\today}

% --------------------
% Abstract
% --------------------
\begin{abstract}
We provide the first non-perturbative lattice determination of the bubble nucleation rate as seeded by topological defects during a first order phase transition. Our case of study is the cubic anisotropy model, which can mimic the Higgs-plus-singlet setup for the electroweak theory, in $d=2+1$ spacetime dimensions, where bubbles are seeded by (line-like) domain walls. We compare the nucleation rate from the lattice with the semi-classical prediction based on the effective field theory living on the domain walls, including for the first time the fluctuation determinant away from spherical symmetry. Our results show very good agreement across all the considered parameter space.
\end{abstract}

\maketitle

% --------------------
% Introduction
% --------------------
\section{Introduction}

First order phase transitions are characterized by the simultaneous presence of two (or more) physically distinct vacua separated by an energy barrier. The false vacuum then decays to the true vacuum via the nucleation of bubbles of the new phase. In a cosmological context, this process is often assumed to be driven by thermal fluctuations in a homogeneous background with uniform probability in space\,\cite{Linde:1980tt,Linde:1981zj}. This bears certain implications for the most likely (critical) bubbles, notably spherical symmetry.

The presence of impurities in the early Universe (for instance, topological defects such as domain walls\,\cite{Blasi:2022woz,Agrawal:2023cgp,Blasi:2023rqi,Bai:2025qch}, strings\,\cite{Steinhardt:1981mm,Steinhardt:1981ec,Yajnik:1986tg,Yajnik:1986wq,Dasgupta:1997kn,Lee:2013zca,Blasi:2024mtc,Chatrchyan:2025uar}, and monopoles\,\cite{Steinhardt:1981mm,Steinhardt:1981ec,Kumar:2009pr,Kumar:2010mv,Agrawal:2022hnf}) can, however, dramatically change this picture by catalyzing bubble nucleation in their vicinity. As a consequence, the nucleation probability is no longer uniform in space but rather enhanced around the defects, and the symmetry of the critical bubble will generally be reduced. 
Cosmological phase transitions can be sources of a stochastic gravitational wave background; see e.g.\,\cite{Caprini:2019egz} for a review. A seeded nucleation will drastically change the distance- and timescales associated with the transition and hence the resulting signal, see  \cite{Witten:1984rs}, and \cite{Blasi:2023rqi,Chatrchyan:2025uar} for a detailed study of the gravitational wave spectrum.

The framework for describing seeded bubble nucleation in quantum and thermal field theory follows essentially the semi-classical methods developed in\,\cite{Coleman:1977py,Callan:1977pt,Linde:1980tt,Linde:1981zj} for the homogeneous case. The probability of seeded nucleation is still dominated by the saddle-point of the action corresponding to the critical bubble, albeit with less symmetry than the usual $\mathrm O(4)$ or $\mathrm O(3)$ solutions. Furthermore, as seeded nucleation is restricted along the sub-manifold where the impurity is located, one has in general a different number of zero modes corresponding to spatial translations compared to the homogeneous case, where nucleation can occur anywhere in space. 

Physics of bubble nucleation can also be studied by means of lattice field theory simulations. For strongly suppressed transitions, as in most cosmologically interesting models, the lattice is used to investigate the configurations near the saddle-point in configuration space. This approach is very similar to that of Langer's classical investigation in \cite{langerStatisticalTheoryDecay1969}. For its lattice field theory analog, see e.g. \cite{Moore:2001vf,Moore:2000jw,Gould:2022ran, Gould:2024chm}. 
The main advantage of this approach is to bypass the approximations that are inevitably introduced in semi-classical calculations and provide an independent, non-perturbative, determination of the bubble nucleation rate. Moreover, the lattice can be used to test the validity of the perturbative calculations against simulation data, see e.g.\,\cite{Gould:2024chm,Niemi:2024axp}.

Alternatively, for sufficiently large nucleation rates, a more intuitive method can be used: assuming that the time evolution of the system can be simulated on a computer, an initial metastable configuration can be evolved in time until a nucleation event happens. Repeated simulations can be performed to collect statistics on e.g. the average lifetime, which easily translates into the nucleation rate. 

We will specify the dynamical evolution of the fields with the stochastic Hamiltonian method \cite{horowitz_stochastic_1985,shijong_canonical_1985}, which is a mixture of Hamiltonian and Langevin evolution with a tunable noise. The final rate will depend on the magnitude of the noise. 
The Hamiltonian evolution has the additional theoretical benefit that it describes the evolution of a quantum scalar theory at high temperatures to leading order in couplings \cite{Bodeker:1996wb,Aarts:1997kp}.
The stochastic Hamiltonian equations are relatively easily simulated on a lattice. For early numerical works in kink-antikink pair production see \cite{soliton_pair_1988,bochkarev_study_1989,alford_thermal_1992} and for nucleation \cite{nucleation_in_1990,alford_metastability_1993,borsanyi_fate_2000}. For more recent works in real-time simulations of thermal false vacuum decay see \cite{pirvu_thermal_2024, batini_real-time_2024, Hirvonen:2025hqn, pirvu_thermal_2026}.

All lattice studies to date have focused on homogeneous phase transitions. It then appears highly desirable to extend this method to phase transitions catalyzed by impurities, which is the scope of this paper. Such a first-principle determination of the bubble nucleation rate will allow us to test the theoretical framework of seeded tunneling and, in particular, the one developed in\,\cite{Blasi:2022woz}. The simulation setup we are going to use is a straightforward generalization of the one employed for homogeneous nucleation: we include a defect in the simulation box at the initial time and monitor the nucleation of bubbles around it via a real-time approach. For mere computational-time convenience, we focus on the cubic anisotropy model in $2+1$ dimensions, where the impurity is a (line-like) domain wall. Our setup  mimics the one considered in\,\cite{Blasi:2022woz,Agrawal:2023cgp,Bai:2025qch} for a seeded electroweak phase transition, and can be straightforwardly generalized to a $3+1$ spacetime.

With the aim of providing a complete theoretical prediction to be tested against simulation data, we will also highlight how the determinant of the fluctuation operator, which contributes to what is usually referred to as the prefactor\,\cite{Callan:1977pt} of the nucleation rate, can be computed for the case of seeded tunneling. In fact, this is a non-trivial task due to the lack of spherical symmetry of the critical bubble which most methods rely on (see e.g.\,\cite{Ekstedt:2023sqc}), which can however be carried out thanks to the domain-wall effective theory introduced in\,\cite{Blasi:2022woz}.

\section{Setup}
\label{sec:setup}
In this paper we are going to be working with the cubic anistotropy model, which has already been the subject of lattice studies of bubble nucleation in the literature\,\cite{Moore:2001vf}. This setup can also describe the essential particle physics content of the extended SM in the $\mathbb{Z}_2$ symmetric limit, where the electroweak phase transition can be first order and seeded by domain walls\,\cite{Blasi:2022woz,Agrawal:2023cgp,Bai:2025qch}. 

We will consider a realization of this scenario in $d$ spatial dimensions, and eventually set $d=2$ for computational convenience. This $d$-dimensional theory is understood as coming from the corresponding thermal field theory in $d+1$ dimensions at finite temperature $T$ after performing standard dimensional reduction.
The model is then described by
\begin{equation}
    Z \propto \int \mathcal{D}\phi_a \, \mathcal{D}\phi_b \,e^{-S[\phi_a,\phi_b]},
\end{equation}
with action 
\begin{equation}
\label{eq:cubicanisotropy}
  S= \frac{1}{T}\int \mathrm{d}^{d}x  \left[ \frac{1}{2} (\partial_i \phi_{\alpha})^{2} - \frac{{m_{\alpha}^2}}{2} \phi_{\alpha}^{2} + \frac{{\lambda^\prime_{\alpha}}}{24} \phi_{\alpha}^{4}  + \frac{\mu^\prime}{4} \phi_{a}^{2}\phi_{b}^{2} \right],
\end{equation}
where $\alpha=a,b$ and $i =1,\dots,d$. Even though not indicated explicitly for notational convenience, the scalar masses as well as the quartics do in general depend on $T$.

Through the field redefinitions $\phi_\alpha \rightarrow \sqrt{T}\,\phi_\alpha$, the dependence on the temperature 
can be reabsorbed in the model parameters such that
\begin{equation}
\label{eq:cubicanisotropy0}
  S= \int \mathrm{d}^{d}x  \left[ \frac{1}{2} (\partial_i \phi_{\alpha})^{2}+ V(\phi_a,\phi_b) \right],
\end{equation}
where
\begin{equation}
    V(\phi_a,\phi_b) = \sum_{\alpha=a,b}\left(- \frac{{m_{\alpha}^2}}{2} \phi_{\alpha}^{2} +\frac{{\lambda_{\alpha}}}{24} \phi_{\alpha}^{4}\right)  + \frac{\mu}{4} \phi_{a}^{2}\phi_{b}^{2},
\end{equation}
and $\lambda_\alpha  = T\, \lambda_\alpha^\prime$ and $\mu = T \,\mu^\prime$.

We will focus on the region of the model parameter space where first $\phi_b$ obtains a non-zero vacuum expectation, $\langle \phi_b \rangle =\pm v_b$ (while $\langle \phi_a \rangle = 0$), leading to the formation of domain walls from the spontaneous $\mathbb{Z}_2$-symmetry breaking. The domain-wall profile can actually be obtained analytically for the $\phi^4$-theory as
\begin{equation}
\label{eq:profile}
    \phi_{b, \,\rm DW}(x) = v_b \,\tanh\left(m_b \, x/\sqrt{2}\right),
\end{equation}
which interpolates between $\phi_b = \pm v_b$ at $x= \pm \infty$.

Later on, the false vacuum, $(\pm v_b,0)$, decays to the true vacuum of the theory at zero temperature, $(\pm v_a,0)$, via the nucleation of bubbles.
The thermal history can then be summarized as:
\begin{equation}
    (\phi_a,\phi_b) \colon (0,0) \xrightarrow{\text{DWs form}} (0,\pm v_b) \xrightarrow{\text{1st order}} (\pm v_a, 0),
\end{equation}
where the second step can in principle proceed by both homogeneous bubble nucleation far from the domain walls, and by seeded nucleation around the domain walls (namely at $x=0$ with reference to \eqref{eq:profile}), albeit with different nucleation probability.

\begin{figure}[htbp]
\includegraphics[]{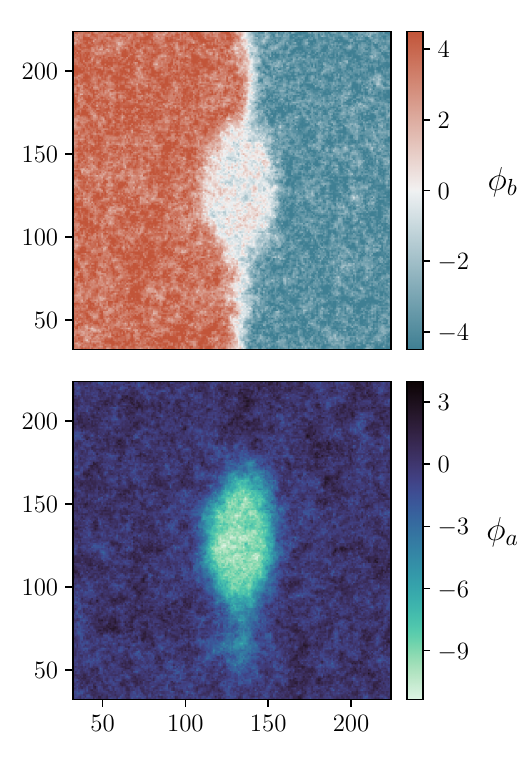}
\caption[]{Snapshot of the two fields a while after the nucleation event. The upper figure shows the wall-field $\phi_b$, while the lower one shows the nucleating field $\phi_a$, here tunneling to the vacuum at $(0,-v_a)$.}
\label{fig:nucleation-image}
\end{figure}

The focus of our work is to study the second step of the phase transition. We will then consider the $(0,\pm v_b)$ vacuum as the initial state, including the presence of domain walls interpolating between $\phi_b=\pm v_b$.
Within our setup, $\phi_a$ plays the role of the Higgs\,\footnote{For the SM Higgs, the $\pm v_a$ vacua would be gauge equivalent.} during the electroweak phase transition, and $\phi_b$ is the BSM scalar singlet. In order to more efficiently simulate the system on the lattice, we will focus on a $d=2$ dimensional realization of this model, where domain walls are line-like defects. To illustrate our setup, we show a simulation snapshot of a nucleation event seeded by a $\phi_b$-domain wall in Fig.\,\ref{fig:nucleation-image}.

As our main goal is to derive a theoretical prediction for the tunneling rate per unit domain-wall length and compare it with our results from the lattice, we need to choose a dynamics for the system. We will consider the fields as classical, coupled to a thermal reservoir. The evolution of the fields is then governed by the stochastic Hamiltonian equations \cite{shijong_canonical_1985}:
\begin{align}
  \frac{\partial \phi_{\alpha}}{\partial t} &= \frac{\delta H}{\delta \pi_{\alpha}} \label{eq:stochastic-hamiltonian-eoms-phi} \\
  \frac{\partial \pi_{\alpha}}{\partial t}  &= - \frac{\delta H}{\delta \phi_{\alpha}} - \gamma \frac{\delta H}{\delta \pi_{\alpha}} + \xi,
  \label{eq:stochastic-hamiltonian-eoms-pi}
\end{align}
where $\xi$ is a Gaussian noise with
\begin{align}
    \langle \xi(x,t) \xi(x',t')\rangle = 2\gamma \delta^{(2)}(x-x')\delta(t-t').
\end{align}
$\gamma$ adjusts the coupling of the system with the thermal bath. Langevin dynamics is obtained in the limit $\gamma \rightarrow \infty$, while $\gamma = 0$ corresponds to the standard non-dissipative Hamiltonian dynamics.

In this case, the prediction for the nucleation rate in the saddle-point approximation factors into two terms: a dynamical part, $A_{\rm dyn}$, and a statistical part, $A_{\rm stat}$. That is,
\begin{equation}
    \Gamma= A_{\rm dyn} \times A_{\rm stat},
\end{equation}
where
\begin{equation}
\label{eq:Adyn}
A_{\rm dyn} = \frac{1}{2\pi} \left( \sqrt{|\lambda_-| + \frac{\gamma^2}{4}} - \frac{\gamma}{2} \right)
\end{equation}
for the stochastic Hamiltonian equations \cite{Langer:1967ax,Langer:1969bc,Csernai:1992tj,Berera:2019uyp,Gould:2021ccf,Ekstedt:2022tqk,Hirvonen:2024rfg,Ekstedt:2023sqc}. $\lambda_-$ is the negative eigenvalue associated with the (unstable) fluctuation around the critical bubble. The statistical factor reads
\begin{equation}
\label{eq:Astat}
    A_{\rm stat} = \left( \frac{B}{2\pi}\right)^{n/2} 
    \left| \frac{\,\text{det}^\prime \,\mathcal{O}[\phi_t]}{\,\text{det}\,\mathcal{O}[\phi_{\rm FV}]} \right|^{-1/2} e^{-(B-B_{\rm FV})},
\end{equation}
where $B$ is the (bounce) action of the critical bubble profile, $\phi_t =\{\phi_{a,t}(\mathbf{x}),\phi_{b,t}(\mathbf{x})\}$, and $B_{\rm FV}$ is the action associated to the false vacuum. The operator $\mathcal{O}$ corresponds to the second derivative of the action, and the determinant comes from integrating over the fluctuations around the tunneling trajectory, $\mathcal{O}[\phi_t]$, or the false vacuum, $\mathcal{O}[\phi_{\rm FV}]$. 
The notation $\mathrm{det}^\prime$ indicates the removal of the $n$ zero modes related to translational invariance. 

For homogeneous nucleation, the false vacuum is given by $\phi_{\rm FV}:(\phi_a=0, \phi_b=\pm v_b)$, with the two possible choices of $\pm v_b$ leading to the same physics. The number of zero modes is $n=d$ corresponding to spatial translations in all possible directions.  The critical bubble profile is spherically symmetric, $\phi_t =\{\phi_{a,t}(|\mathbf{x}|),\phi_{b,t}(|\mathbf{x}|)\}$. As a consequence, the fluctuation operator around the critical bubble, $\mathcal{O}[\phi_t]$, is spherically symmetric as well. While still challenging in general, the evaluation of the determinant can then be tackled numerically by taking advantage of the spherical symmetry, see\,\cite{Ekstedt:2023sqc}.
The negative eigenvalue, $\lambda_-$, can also be readily computed numerically with the tunneling profile at hand.

The type of seeded tunneling around the domain-wall impurity shown in figure\,\ref{fig:nucleation-image} introduces some difficulties for the evaluation of $A_{\rm stat}$ in \eqref{eq:Astat} when compared to the homogeneous case. First of all, the false vacuum is no longer constant in space but rather contains a space-dependent domain-wall configuration as given in \eqref{eq:profile}. This also implies that the critical bubble is no longer spherical, making it technically more difficult to determine its profile and bounce action. Moreover, the evaluation of the fluctuation determinants is severely complicated due to the lack of spherical symmetry. The identification of the zero modes is less obvious as well, even though one expects that only spatial translations parallel to the domain wall should survive.

In the next section, we shall see how these issues can be circumvented by working within the domain-wall effective field theory developed in\,\cite{Blasi:2022woz}.

\section{Prediction for the seeded nucleation rate}

Seeded tunneling can be described as a homogeneous process taking place along the catalyzing defect. This allows to apply all the methods and simplifications developed for standard homogeneous nucleation.

In the domain-wall case of interest, one then constructs
a $d-1$ effective field theory (EFT) by employing a Kaluza--Klein (KK) decomposition of the fields $\phi_{a,b}$ around the domain-wall background: 
\begin{equation}
    \phi_b = \sum_n\phi_b^{(n)}(y) \varphi_b^{(n)}(x) + \phi_{b, \,\rm DW}(x),
\end{equation}
and
\begin{equation}
    \phi_a = \sum_n \phi_a^{(n)}(y) \varphi_a^{(n)}(x),
\end{equation}
where the domain-wall profile is given in \eqref{eq:profile}, and $y$ is the coordinate on the domain wall. The profiles $\varphi_{a,b}(x)$ can be obtained by solving the associated eigenvalue problem.
Such expansion can be inserted into the action \eqref{eq:cubicanisotropy0} and, upon performing the integration along the $x$ direction orthogonal to the domain wall, one obtains a theory for the $\phi^n_{a,b}(y)$ modes living on the wall. 

By exploiting some hierarchy in the mass spectrum of the theory, one may retain only the lightest states among $\phi_a^{(n)}$ and $\phi_b^{(n)}$, indicated here by $\phi_a^0$ and $\phi_b^0$, respectively, as dynamical fields.
This leads to the following action in the EFT\cite{Blasi:2022woz}:
\begin{equation}
\label{eq:SDW}
  S_{\rm EFT} = \int_{-\infty}^\infty \mathrm{d}y \left[\frac{1}{2} \left(\frac{{\rm d}\phi_{\alpha}^0}{{\rm d}y} \right)^{2} + \frac{{\tilde m_{\alpha}^2}}{2} \phi_{\alpha}^{0\,2} + \tilde V(\phi_a^0,\phi_b^0)\right],
\end{equation}
where $\alpha=a,b$, and we denoted by $\tilde m_\alpha$ the mass of the $\phi_{a,b}^0$ modes.
These are related to the original model parameters in \eqref{eq:cubicanisotropy0} as
\begin{equation}
    \tilde m^2_a = -m^2_a + \frac{1}{2} p m^2_b, \quad p= \frac{1}{2} \left(\sqrt{1 + \frac{24 \mu}{\lambda_b}}-1 \right),
\end{equation}
and
\begin{equation}
\label{eq:mtildeb}
    \tilde m^2_b = \frac{3}{2} m^2_b.
\end{equation}
The other KK states have masses $m^2_{\rm KK} \gtrsim m^2_b$, and their effect is captured by higher-dimensional operators in the effective potential $\tilde V(\phi_a^0,\phi_b^0)$.
This contains interactions mediated by the tree--level exchange of heavy KK resonances up to $\mathcal{O}(m_{\rm KK}^{-4})$ accuracy, and is therefore an 8th-order polynomial. (Increasing the accuracy beyond this level is challenging due to the large number of possible tree-level exchanges, and would also require to include derivative operators). Notice that we kept $\phi_b^0$ as dynamical even though $\tilde m^2_b \sim m^2_{\rm KK}$. As we shall see, the final result for the bounce action does not strongly depend on this choice, which is consistent with our EFT approach.

Depending on the portal coupling $\mu$, the mode $\phi_a^0$ can have a positive mass\,\footnote{The case of $\tilde m^2_a <0$ would signal a classical instability of the domain wall, which would then dissociate without encountering any energy barrier.}. This then controls the leading-order barrier for seeded nucleation, which in the EFT is described by the following transition: 
\begin{equation}
(\phi_a^0,\phi_b^0):(0,0) \rightarrow (\langle \phi_a^0 \rangle,\langle \phi_b^0 \rangle).
\end{equation}
Notice that this process corresponds to standard homogeneous nucleation, as seeded bubbles have the same probability of being nucleated at any point along the domain wall. 

For the EFT to work, one needs $\tilde m_a^2 \ll \tilde m _b^2$, which can be satisfied in certain regions of the parameter space. In the following, we shall assume that this it the case. The rate for the seeded tunneling of interest can then be computed straightforwardly according to the general framework presented in section \ref{sec:setup} within the EFT in \eqref{eq:SDW}. In particular, the statistical factor for seeded tunneling is given by
\begin{equation}
\label{eq:Astatseed}
    A_{\rm stat}^{s} = \left( \frac{B_s}{2\pi}\right)^{1/2} 
    \left| \frac{\,\text{det}^\prime \,\mathcal{O}_s[\phi_t^0]}{\,\text{det}\,\mathcal{O}_s[\phi_{\rm FV}]} \right|^{-1/2} e^{-(B_s-B_{\rm FV})}.
\end{equation}
The action of the seeded critical bubble, $B_s$, can be obtained as
\begin{equation}
\label{eq:Bs}
    B_s = 2 \int_0^{\bar \phi} \text{d}\phi_t \sqrt{2 \tilde V(\phi^0_t)},
\end{equation}
where $\phi^0_t$ is the tunneling direction in the two-field space given by $(\phi_a^0,\phi_b^0)$, and $\bar \phi$ corresponds to the point where $\tilde V(\bar \phi)=0$ along this trajectory\,\footnote{The effective potential $\tilde V$ is such that it vanishes at the false vacuum, $(\phi_a^0=0,\phi_b^0=0)$.}. This expression for $B_s$ is particularly simple as we are working with a line-like defect, and the factor of two comes from the symmetry around $y=0$. Notice that we have a single zero mode ($n=1$) in \eqref{eq:Bs}, as this corresponds to translations of the critical bubble along the domain wall. Within our conventions, one has $B_{\rm FV}=0$.

As for the functional determinants in \eqref{eq:Astatseed}, they can actually be computed analytically by introducing a further simplification of the EFT in \eqref{eq:SDW}. First of all, we integrate out $\phi_b^0$ at tree level to obtain a one-field tunneling problem for $\phi_a^0$. This is justified as $\phi_b^0$ is rather heavy, as discussed below \eqref{eq:mtildeb}. One then has:
\begin{equation}
    \label{eq:SDW1field}
  S_{\rm EFT} \simeq \int \mathrm{d}y \left[ \frac{1}{2} \left(\frac{{\rm d}\phi_{a}^0}{{\rm d}y} \right)^{2} + \frac{{\tilde m_{a}^2}}{2} \phi_{a}^{0\,2} - \frac{\lambda}{4}\phi_a^{0\,4} + \dots \right],
\end{equation}
where $\lambda >0$ is the resulting effective quartic coupling. The dots indicate further terms that are not essential for the tunneling problem and will be neglected. In this approximation, the bounce solution satisfies
\begin{equation}
    -\phi_t^{\prime\prime}(y) + \tilde m^2_a \phi_t(y) -\lambda \phi_t^3(y) = 0, \quad \phi_t(\pm \infty)=0.
\end{equation}
The fluctuation operator is then given by
\begin{equation}
\label{eq:Os}
    \mathcal{O}_s[\phi_t] = - \frac{\mathrm d^2}{\mathrm dy^2} + \tilde m_a^2 - 3 \lambda \phi_t^2(y).
\end{equation}
At $y=0$ one has $\phi^\prime_t(0)=0$ and $\phi_t^2(0)= 2 \tilde m_a^2/\lambda$. This leads to the observation that $\mathcal{O}_s[\phi_t]$ can be approximated by a Pöschl-Teller potential of the form
\begin{equation}
\label{eq:Obk2}
\mathcal{O}_{\rm PT}\equiv - \frac{\mathrm d^2}{\mathrm dy^2} + \tilde m^2_a \left[1 + k^2 - \frac{j(j+1)}{\text{cosh}^2(\tilde m_a y)} \right]
\end{equation}
with $j=2$. One can check that the spectrum of $\mathcal{O}_{\rm PT}$ has exactly two bound states with energy levels $\lambda_l = \tilde m^2_a(1-l^2)$ with $l=1,2$, yielding a zero mode with $\lambda_0=0$ and one mode with negative eigenvalue 
\begin{equation}
\lambda_- = - 3 \tilde m_a^2,
\end{equation}
as expected from our tunneling problem. The operator related to fluctuations in the false vacuum, $\mathcal{O}_s[\phi_{\rm FV}]$, is instead simply given by \eqref{eq:Obk2} with $j=0$.
In order to later on extract the zero mode from the determinant around the bounce solution and regularize our procedure, we have included the $k^2$ term in \eqref{eq:Obk2}, which we will eventually send to zero. 

The ratio of the determinants in \eqref{eq:Astatseed} can then be approximated by:
\begin{equation}
 \lim_{k\to0} \left| \frac{1}{\tilde m_a^2 k^2}\frac{\,\text{det}\,\mathcal{O}_{\rm PT}|_{j=2}}{\,\text{det}\,\mathcal{O}_{\rm PT}|_{j=0}} \right|^{-1/2} = 2 \sqrt{3} \tilde m_a,
\end{equation}
where this result can be obtained with the Gelfand-Yaglom theorem, as well as from a direct computation of the phase shifts and bound-state energy levels\,\cite{Dunne:2007rt}.

The full rate for seeded bubble nucleation per unit domain wall length, including the fluctuation determinant can then be approximated by:
\begin{equation}
\label{eq:rate}
    \Gamma^s = A_{\rm dyn}^s \times A_{\rm stat}^s  \simeq 2 \times \frac{3}{\pi} \tilde m_a^2 \left( \frac{B_s}{2\pi}\right)^{1/2} e^{-B_s} f(\gamma/\tilde m_a),
\end{equation}
where
\begin{equation}
    f(t) = \sqrt{1+t^2/12}-t/\sqrt{12},
\end{equation}
and the factor of two takes into account the $\mathbb{Z}_2$ symmetry for the field $\phi_a$, which can tunnel to either of the $\pm v_a$ minima with the same probability. $A^s_{\rm dyn}$ is given by \eqref{eq:Adyn} with $\lambda_- = - 3 \tilde m_a^2$, as discussed above. 

Overall, we expect the P\"osch-Teller approximation to introduce only a factor of $\mathcal{O}(1)$ uncertainty in the theoretical prediction of $\Gamma^s$. Let us also notice that the prefactor that we obtained differs significantly from what one could guess based on dimensional analysis. In particular, this turns out to be controlled by $\tilde m_a^2$, which can be order-of-magnitude smaller than all the other scales in the problem such as the domain wall tension, which in this case is $\sigma_{\rm DW} \sim m_b^2 \gg \tilde m_a^2$.

\subsection{Benchmark points}
\label{sec:ben}

With $d = 2$ spatial dimensions the fields are dimensionless, and the mass dimensions of the couplings are given by
\begin{equation}
    \big[m_{\alpha}^2\big] = \big[\lambda_{\alpha}\big] = \big[ \mu \big] = 2.
\end{equation}
\emph{For the rest of the paper we implicitly write all quantities in units of an arbitrary mass scale $\scale$}. Then the benchmark points relevant for our lattice study are given by the masses
\begin{equation}
    m_a^2 = 0.09, \quad m_b^2 = 0.226,
\end{equation}
and quartic couplings
\begin{equation}
\label{eq:benchmarkpoint}
\lambda_a = 0.006, \quad \lambda_b = 0.096, \quad \mu = 0.026 \,(0.028).
\end{equation}
The masses of the states in the domain-wall EFT read
\begin{equation}
\tilde m_a^2 = 0.008 \,(0.013), \quad \tilde m_b^2 = 0.34,
\end{equation}
so that the condition $\tilde m_a^2/\tilde m_b^2 \ll 1$ is satisfied. The value of $\tilde m_a^2$ in parenthesis refers to $\mu = 0.028$.
The bounce action, $B_s$, evaluated with different accuracy in the domain-wall EFT are reported in Table \,\ref{table:theory} for the two values of $\mu$ considered in \eqref{eq:benchmarkpoint}. The different orders of accuracy correspond to higher-order operators that we include in the effective potential $\tilde V(\phi_a^0,\phi_b^0)$ in \eqref{eq:SDW}, ranging from neglecting altogether the KK excitations to keeping terms up to $\mathcal{O}(m_{\rm KK})^{-4}$. We also indicate in parenthesis the value of the seeded bounce action obtained by further integrating out the $\phi_b^0$ mode.
The knowledge of $B_s$ together with $\tilde m_a^2$ allows then a straightforward prediction for the tunneling rate per unit wall length via \eqref{eq:rate}, once the damping $\gamma$ is specified.

\setlength{\tabcolsep}{6pt} % Default value: 6pt
\begin{table}
  \begin{tabular}{@{}ccccc@{}}
    \toprule
    & & \multicolumn{3}{c} {$B_s$}  \\
    \cmidrule(){3-5}
        $\mu$ & $\tilde m_a^2$ &  No KK & $\mathcal{O}(1/m^2_{\rm KK})$ & $\mathcal{O}(1/m^4_{\rm KK})$ \\
        \cmidrule(r){1-2}\cmidrule(){3-5}
    0.026 & 0.008 & 6.05 (5.86) & 5.23 (5.07) & 5.27 (5.11) \\ 
    0.028 & 0.013 & 12.0 (11.4) & 9.67 (9.25)& 9.74 (9.30) \\
    \bottomrule
  \end{tabular}
  \caption{Bounce action of the critical bubble, $B_s$, for different values of the portal coupling $\mu$ (the other model parameters are specified in Sec.\,\ref{sec:ben}), calculated within the DW EFT at different orders of accuracy: neglecting KK excitations (first column), at $\mathcal{O}(1/m^2_{\rm KK})$ (second column), and at $\mathcal{O}(1/m^4_{\rm KK})$ (third column). The value quoted in parenthesis is obtained by further integrating out $\phi_b^0$, leading to a one--field tunneling problem for $\phi_a^0$. For each value of $\mu$, the various predictions indicate the degree of theoretical uncertainty for $B_s$. We also report the derived quantity $\tilde m_a^2$ for each $\mu$.}
  \label{table:theory}
\end{table}

Let us also notice that by rescaling the quartic couplings, $\lambda_{\alpha}$ and $\mu$, by a common factor $x < 1$, we can effectively describe a decrease in temperature and therefore obtain a more suppressed nucleation rate. Within our approximation for the decay rate, this will only affect the value of $B_s$, which will change to $B_s/x$ when evaluating the rate in \eqref{eq:rate}.

Given the exponential decay law in time for the false vacuum, the average life-time of the false vacuum is
\begin{equation}
    \tau = \frac{1}{\Gamma^s \cdot L_{\rm w}},
\end{equation}
where $\Lwall$ is the total length of the domain wall in the simulation box.

Far from the domain walls, the nucleation of spherical bubbles can of course still take place. However, this process is predicted to be extremely slow compared to seeded tunneling for the benchmark points under consideration. In particular, the bounce action far from the walls turns out to be $B \gtrsim 100$. We thus expect to see no such event in our simulation box, and the phase transition to complete thanks to seeded bubbles.

\section{Lattice implementation}

The simulation was written using the lattice programming framework HILA \cite{HILALatticeSimulation}. The system is simulated on a two-dimensional rectilinear lattice with $N_y$ sites along the domain wall and $N_x$ to the perpedicular direction. The lattice is periodic in the $y$-direction, along the domain wall, and antiperiodic in the $x$-direction. The latter boundary condition forces the appearance of a domain wall. The physical dimensions are simply $(L_x, L_y) = (a N_x, a N_y)$. Where appropriate, we use $\Lwall$ instead of $L_y$. The scalar fields are defined on the lattice sites and the Hamiltonian is discretized using finite differences. This gives the lattice Hamiltonian as
\begin{align}
\label{eq:HL}
  H_{\mathrm L} =  \sum_{n} \sum_{\alpha =a,b} & a^{2} \bigg( \frac{\pi_{\alpha}^{2}}{2} + \frac{Z_{\phi}}{2} (\phi_{\alpha}\nabla_{\mathrm L}^{2}\phi_{\alpha}) \\
                        & + \frac{{Z_{m_{\alpha}}(m_{\alpha}^2 + \delta m_{\alpha}^{2})}}{2} \phi_{\alpha}^{2} \\
                        &  + \frac{{\lambda_{\alpha} + \delta \lambda_{\alpha}}}{24} \phi_{\alpha}^{4} \bigg) + \frac{\mu + \delta \mu}{4} \phi_{a}^{2}\phi_{b}^{2},
\end{align}
where the field variables are implicitly functions of the lattice site $n$, $a$ is the lattice spacing and $\nabla_{\mathrm L}^{2}$ the $\mathcal O(a^{2})$ accurate lattice laplacian
\begin{align}
  \nabla_{\mathrm L}^{2}\phi(n) &= \frac{1}{a^2} \sum_{i}\bigg[\frac{4}{3} \left( \phi(n + i) + \phi(n-i) \right) \\
  & - \frac{1}{12} \left( \phi(n + 2i) + \phi(n - 2i) \right) - \frac{5}{2} \phi(n) \bigg].
\end{align}
The lattice counterterms are discussed in App.\,\ref{sec:counter}.

This is now a coupled system of $4 N_{x} N_{y}$ scalar functions governed by the stochastic Hamiltonian equations \ref{eq:stochastic-hamiltonian-eoms-phi} and \ref{eq:stochastic-hamiltonian-eoms-pi}. We evolve the system numerically using a fourth order symplectic Yoshida integrator \cite{yoshida_construction_1990}. We take the stochastic term $\xi$ into account by updating the momenta every $\Delta t \ll a$ as
\begin{equation}
\pi_{\alpha}(x,t+0) = (1 - \epsilon^{2})^{1/2}\pi_{\alpha}(x,t-0) + \epsilon \eta_{\alpha}(x,t),
\end{equation}
where $\epsilon^{2} \coloneqq 1 - \exp(- 2 \gamma \delta t)$ and $\eta_{\alpha}$ is drawn from $\mathcal N (0,1)$ \cite{Moore:2001vf}. By a direct computation one finds
\begin{equation}
\langle \pi_{\alpha}(x,t-0)^{2} \rangle  = \langle \pi_{\alpha}(x,t+0)^{2} \rangle,
\end{equation}
meaning that a Gibbs distribution is preserved. Since the other part of the time evolution damps the momenta, the distribution of $\pi_{\alpha}(x,t)$ settles to that of $\eta_{\alpha}$ over time.

The simulation proceeds as follows: First, the system is initialized such that the wall-field $\phi_{b}$ is set to one vacuum on the left half of the lattice and in the other on the right. This creates a sudden jump in the middle, but due to the anti-periodic boundary conditions, the jump is not \enquote{visible} when wrapping around the lattice. The field $\phi_{a}$ is initialized to zero.

\begin{figure}[htbp]
\centerline{\includegraphics[]{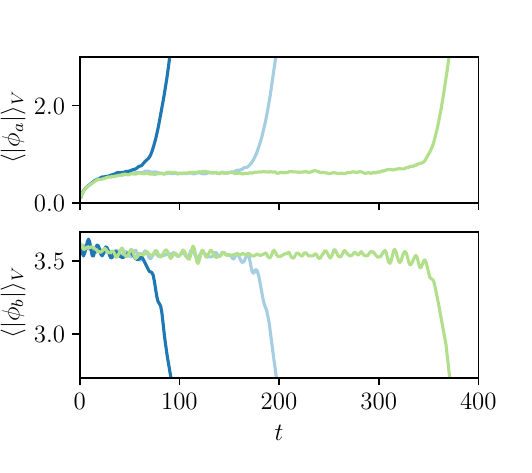}}
\caption[]{Some example trajectories from the simulations at the reference point with the damping factor $\gamma = 0.05$. We graph the volume average of the absolute value of the two fields.}
\label{fig:trajectories}
\end{figure}

The wall-field is then evolved in a static background with $\phi_{a} = 0$ to relax $\phi_{b}$ into a correctly shaped domain wall. Once the wall-field has been relaxed, both fields are evolved until the volume average of $|\phi_{a}|$ reaches some reasonable threshold value, indicating the formation of a bubble. We graph some of the trajectories in Fig. \ref{fig:trajectories}.

\begin{figure}[t]
\includegraphics[]{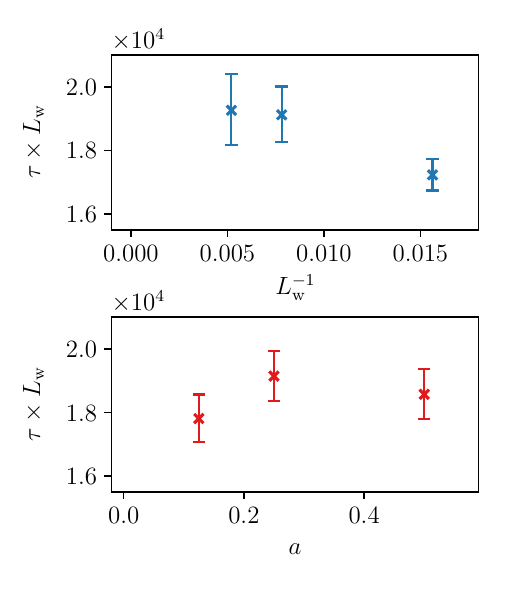}
\caption[]{Lattice dependency of the obtained nucleation times at the reference parameter point. In the upper plot we use $a = 0.5$ and vary the length of the wall $\Lwall \in \{ 64, 128, 196 \}$. In the lower plot we use lattice with the physical dimensions $(L_x, \Lwall) = (64, 128)$ at different values of lattice spacing $a$.}
\label{fig:lattice-dependency}
\end{figure}

\begin{figure}[t]
\includegraphics[]{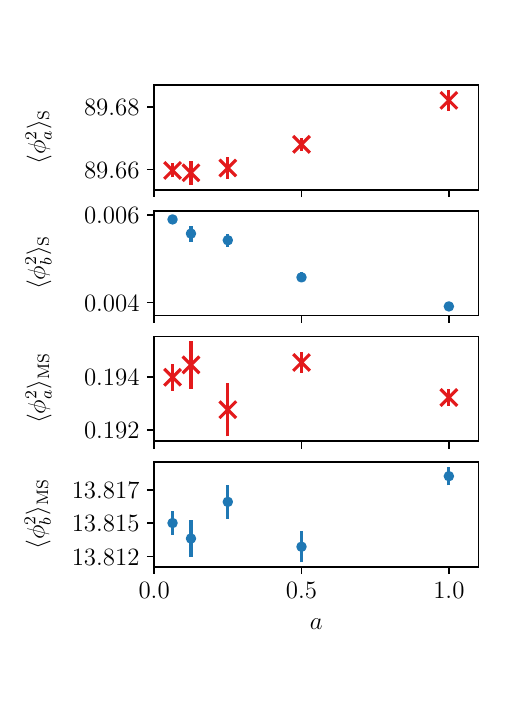}
\caption[]{Lattice spacing dependency of the condensates at the reference parameter point. Here the system does not contain a wall and the boundary conditions are periodic. The physical dimensions of the lattice are $(L_x, L_y) = (64, 64)$. In the metastable phase the dependency is very mild. In the stable phase $\phi_b$ is quite strongly affected by the large vacuum expectation value of $\phi_a^2$ through the mixing interaction.}
\label{fig:lattice-spacing-condensate}
\end{figure}

We tested the lattice spacing and volume dependency of the average nucleation time $\tau$ for the parameters of the benchmark point in Fig. \ref{fig:lattice-dependency}. To ascertain that our renormalization prescription is working, we also measured the values of the scalar condensates and present the results in Fig. \ref{fig:lattice-spacing-condensate}. In the true vacuum the large vacuum expectation value of $\phi_a^2$ makes $\phi_b$ very heavy, leading to an increased lattice spacing dependency. In light of the relatively mild lattice spacing dependence of the measured nucleation times and the number of parameter points investigated, we have elected to perform all simulations with $a = 0.5$ and $(L_x, \Lwall) = (64, 128)$.

\subsection{Extracting the rate}
\begin{figure}[htbp]
\includegraphics[]{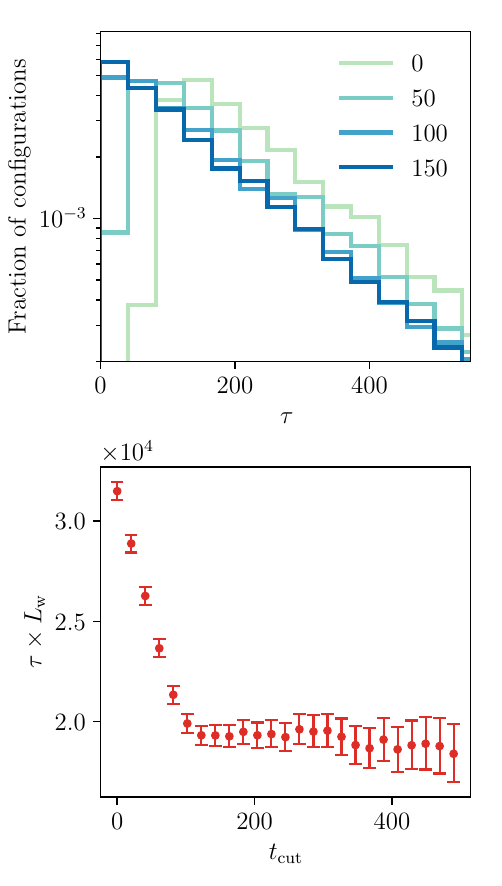}
\caption{The histograms contain the distributions of the normalized nucleation times as the cutoff time is varied. The scatter plot shows the normalized nucleation times extracted from distributions at different cutoff values. For these parameters the rate appears to stabilize at $t \approx 120$, which is not completely obvious from looking at the trajectories of $\langle \left| \phi_{a} \right|\rangle_{V}$ in Fig. \ref{fig:trajectories}. }
\label{fig:cutoff-example}
\end{figure}

From the preceding initialization description it is obvious that the fields at time $t=0$ are not sampled from the thermal distribution and the nucleation rate can not be expected to be constant in time. Of course, due to the stochastic evolution, the fields thermalize over some characteristic time scale $\mathcal T$. To control the effect of the unphysical initial conditions, we analyze the nucleation times using different cutoff times. That is, we take the full set of nucleation times and set a new $t = 0$ at some $\tcut$, discarding all runs where the nucleation occurred before $\tcut$. By varying $\tcut$ (which is easily done in post-processing) we obtain different sets of nucleation times. If we increase $\tcut$, the dependence on the initial conditions should weaken. We then expect that the rates should stabilize as $\tcut \rightarrow \infty$. An example is shown in \ref{fig:cutoff-example}.

We can only analyze a finite number of nucleation events and as the nucleation times are exponentially distributed, this method is practical only when $\mathcal T$ is not much longer than the average nucleation time. However, if $\mathcal T$ were much larger than the thermalization time, a thermal initial distribution might not make physical sense anyways. For all parameter points investigated the assumption remains valid and the cutoff method does not necessitate excessive discarding of nucleation events. 

There are many ways to obtain the average nucleation time from the generated distributions. If we assume that the lifetimes are exponentially distributed, $p(t) \propto \exp(- t / \tau)$, the simplest way to extract $\tau$ is to simply take the mean of the nucleation times $\{t_{i}\}$. Alternatively, we can fit the distribution $p(t)$ of the exponential form to a histogram of the nucleation times. The two methods yield compatible results. We only use the first method in this work.

Having obtained $\tau$ for each cutoff value of interest, there is no straightforward way to choose the cutoff values for the final estimate. The data points in Fig. \ref{fig:cutoff-example} are highly dependent, since they are all derived from the same initial set of nucleation times. For this work we are satisfied with a constant fit to the interval $\tcut \in [200, 400]$.

The error estimates are based on bootstrap resampling the initial set of nucleation times. For each resample we perform the previous analysis and obtain a set of bootstrap estimates for $\tau$. The result is simply the mean of the bootstrap estimates while the error bars denote their 5th and 95th percentiles.

\section{Results}

We give all lattice quantities as dimensionless. Physical units can be recovered by the appropriate multiplication by the scale $\scale$. For these comparisons the scale doesn't have to be specified. Simulations were at the reference parameter point with details given in Table \ref{table:run-details}. In all simulations we set lattice spacing to $a = 0.5$, the time step to $\Delta t = 0.05 a$ and use an elongated lattice with  $(N_x, N_y) = (128, 256)$, where the domain wall runs along the longer side. Different values of the pure number $x$ correspond to transforming the quartic couplings via the scaling $\{\lambda_{\alpha}, \mu\} \rightarrow \{x\lambda_{\alpha}, x\mu\}$. The corresponding change in the nucleation rate is discussed in section \ref{sec:ben}.

\setlength{\tabcolsep}{10pt} % Default value: 6pt
\begin{table}[]
  \sisetup{table-format=2.2+-1.2 }
  \vspace{7mm}
  \begin{tabular}{@{}cccS[separate-uncertainty, table-text-alignment = center]c@{}}
    \toprule
        {$\mu$} & {$\gamma$} & {$x$} & {$\tau \times L_{\mathrm w} \times 10^{-4}$} & {$N_{\mathrm{nucl.}}$} \\
        \midrule
        0.026 &            0.05 &             1.0 &                     1.92 \pm 0.08 &           5000 \\ 
        &                 &             0.9 &                     2.81 \pm 0.09 &           5000 \\ 
        &                 &             0.8 &                     5.15 \pm 0.15 &           5000 \\ 
        &                 &             0.7 &                    11.22 \pm 0.37 &           3000 \\ 
        &                 &             0.6 &                    31.42 \pm 1.01 &           3000 \\ 
        \midrule
        &            0.10 &             1.0 &                     2.32 \pm 0.09 &           5000 \\ 
        &                 &             0.9 &                     3.51 \pm 0.11 &           5000 \\ 
        &                 &             0.8 &                     6.08 \pm 0.17 &           5000 \\ 
        &                 &             0.7 &                    13.59 \pm 0.44 &           3000 \\ 
        &                 &             0.6 &                    36.78 \pm 1.13 &           3000 \\ 
        \midrule
        0.028 &            0.05 &             1.1 &                    18.46 \pm 0.75 &           2000 \\ 
        &                 &             1.0 &                    39.46 \pm 1.59 &           2000 \\ 
        &                 &             0.9 &                   102.25 \pm 3.79 &           2000 \\ 
        \midrule
        &            0.10 &             1.1 &                    20.61 \pm 1.07 &           1200 \\ 
        &                 &             1.0 &                    45.05 \pm 2.23 &           1200 \\ 
        &                 &             0.9 &                   119.87 \pm 5.95 &           1200 \\ 
        \bottomrule
  \end{tabular}
  \caption[]{Results for the normalized average nucleation times for all points in Fig. \ref{fig:rate-result}. Since the error estimates from the bootstrap are very close to symmetric, we only show the larger error in the table.}
  \label{table:run-details}
\end{table}

\begin{widetext}
  
\begin{figure}[htbp]
  \centerline{\includegraphics[]{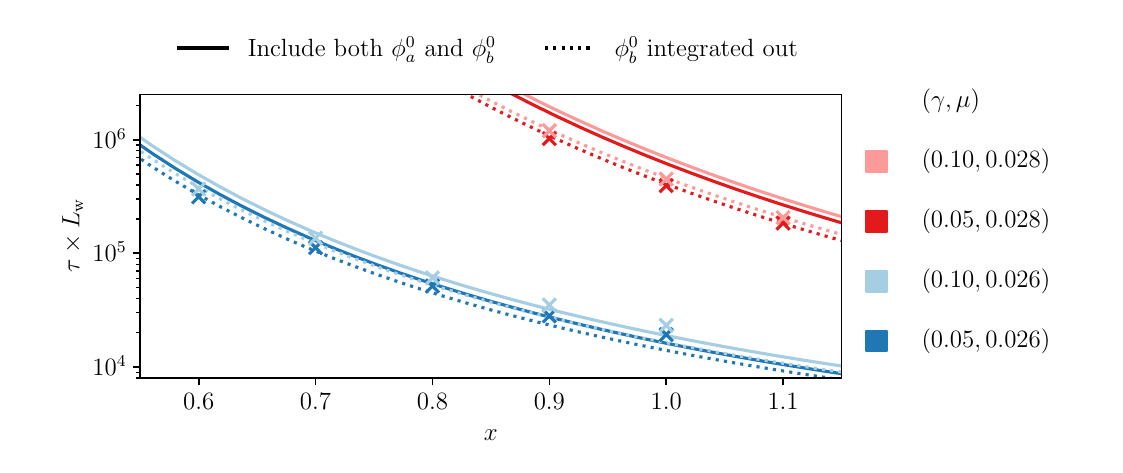}}
\caption{Average configuration lifetime for different values of the quartic scaling factor $x$ for two different values of $\mu$ and $\gamma$. The different lines correspond to different semianalytic estimates for the bounce action $B_s$. The solid line is the $\mathcal O(1 / m_{\mathrm{KK}}^{4})$ estimate in \autoref{table:theory} ($B_s= 5.27$ for $\mu = 0.026$, and $B_s= 9.74$ for $\mu = 0.028$) while the dotted line uses the corresponding parenthesized value ($B_s = 5.11$ for $\mu=0.026$, and $B_s= 9.30$ for $\mu = 0.028$).}
\label{fig:rate-result}
\end{figure}

\end{widetext}

The obtained nucleation time estimates are presented in Fig. \ref{fig:rate-result} along with the semianalytic result. Numerical values are tabulated in Table \ref{table:run-details}. We see that the semianalytic estimates agree well with the simulations. It seems that the simulation results are closer to the more drastic approximation, where only $\phi_a^0$ remains dynamical for the bounce action computation. We will not attribute any special significance to this as the reason we show both semianalytic estimates is only to gauge the uncertainty related to our approximations.

% --------------------
% Conclusions
% --------------------
\section{Conclusions}
We have investigated domain-wall-seeded first order phase transitions in the context of a classical field coupled to a thermal reservoir. We have calculated an estimate for the average lifetime of the metastable phase using a Kaluza-Klein decomposition, thus translating the problem back to that of homogeneous nucleation. We have compared the semianalytic estimate to numerical simulations of the original two-dimensional system and found
very good agreement across the parameter range investigated.

The lattice analysis is in that sense incomplete that we did not perform a detailed continuum or infinite volume extrapolation for the average nucleation times. This should be done to obtain precision estimates but is perhaps unwarranted for our toy model. For very large volumes the scenario of multiple bubbles becomes likely and the method used here becomes somewhat impractical since the cutoff analysis would require us to discard more runs. In addition, the lattice volume averages become noisier indicators as the volume of metastable phase is increased. The latter problem is likely solvable by constructing more refined measurables.

In addition to the real-time approach taken in this work, testing the non-perturbative saddle-point approximation of \cite{Moore:2000jw} would also be interesting. For simplicity such a comparison might initially be made in the scenario of homogeneous nucleation. We do intend to explore this latter method in the case of seeded nucleation in the more physically motivated case of the real scalar singlet extended Standard Model. Since the necessary matching rules to a three-dimensional EFT have already been computed in \cite{Niemi:2024axp}, the system is very amenable to a lattice study.

Let us also comment on how the semianalytic method based on the domain-wall EFT adopted in this paper can be extended in $d=3$ dimensions, where the domain wall is a two-dimensional surface. In this case, the seeded bounce action will not have a simple expression as in \eqref{eq:Bs}, but rather it needs to be obtained by  shooting or path-deformation algorithms as implemented e.g. in \texttt{CosmoTransitions}\,\cite{Wainwright:2011kj}. Moreover, the fluctuation operator will not involve a single spatial dimension as in \eqref{eq:Os}. In this case, we do not expect to find an analytical approximation for the determinant. Nevertheless, we could still evaluate the required determinants with the numerical methods developed in \,\cite{Ekstedt:2023sqc}. Therefore, while more challenging from a computational point of view, we expect to successfully apply this method also in $d=3$.

Finally, let us remark that when the domain wall EFT is outside of its range of validity
a systematic method for evaluating the statistical factor away from spherical symmetry remains an open question.

% --------------------
% Acknowledgments
% --------------------
\begin{acknowledgments}
SB is supported by the Deutsche Forschungsgemeinschaft under Germany’s Excellence Strategy---EXC 2121 ``Quantum Universe"---390833306.  JH and KR are supported by the European Research Council grant ``CoCoS'', number 101142449, and the Research Council of Finland grant number 354572.
\end{acknowledgments}

\newpage
% --------------------
% Appedices
% --------------------
\appendix
\begin{widetext}
\section{Counterterms for the lattice action}
\label{sec:counter}
The discretized lagrangian is written as
\begin{equation}
  \mathcal L_{\mathrm L} =\sum\limits_{n} a^2 \left[ \sum_{\alpha = a,b}  \bigg(\frac{Z_{\phi}}{2} (\phi_{\alpha}\nabla_{\mathrm L}^{2}\phi_{\alpha})
                         + \frac{{Z_{m_{\alpha}}(m_{\alpha}^2 + \delta m_{\alpha}^{2})}}{2} \phi_{\alpha}^{2}
                          + \frac{{\lambda_{\alpha} + \delta \lambda_{\alpha}}}{24} \phi_{\alpha}^{4} \bigg) + \frac{\mu + \delta \mu}{4} \phi_{a}^{2}\phi_{b}^{2} \right],
\end{equation}
\end{widetext}
where $a$ is the lattice spacing and the fields are implicitly functions of the lattice site $n$. The couplings have their renormalized values that are \emph{a priori unrelated to the continuum values}.

The prefactor $a^2$ can be absorbed into the couplings by defining the dimensionless quantities
\begin{equation}
\mlat_{\alpha}^{2} \coloneqq a^2m_{\alpha}^2, \qquad \llat_{\alpha} \coloneqq a^2 \lambda_{\alpha}, \qquad \mulat \coloneqq a^2 \mu.
\end{equation}
Since the two-dimensional theory is super-renormalizable, the lattice couplings go to zero in the continuum limit. In perturbation theory only the one-loop contribution to the two-point function is divergent, meaning that the quartic counterterms can be set to zero. Futhermore, the finite corrections to the quartic lattice couplings and multiplicative renormalization factors $Z_{i}$ are suppressed by higher powers of the vanishing lattice couplings $\mulat$ and $\llat_\alpha$. For what follows, we now set $m_{\alpha}^2$ to the bare continuum value (any other choice could well be absorbed into $\delta m_{\alpha}^2$).

For the lattice theory to match the continuum theory, the lattice couplings must be chosen such that physical quantities are reproduced. For the quartic couplings super-renormalizability guarantees that the value of the four-point function tends to the value of the corresponding lattice coupling as $a \rightarrow 0$. This means that the quartic lattice couplings can be set to the renormalized continuum values $\lambda_{\alpha}^{\cont}$ and $\mu^{\cont}$ and the corresponding counterterms to zero.

We will now compute the necessary correction terms for the two-point function. For our euclidean lattice action the Feynman rules read
\begin{figure}[htbp]
  \begin{tikzpicture}
    \def\len{0.7}
    \def\dist{2.5}
    \def\eqdist{1.0}
    
    \coordinate (A) at (0,0);
    \draw[dashed] (A) -- ++(-\len,-\len);
    \draw[solid] (A) -- ++(\len,\len);
    \draw[dashed] (A) -- ++(-\len,\len);
    \draw[solid] (A) -- ++(\len,-\len);
    \node[] at ($ (A) + (\eqdist,0) $) {$=-\mulat$};
    
    \coordinate (A) at (\dist,0);
    \draw[solid] (A) -- ++(-\len,-\len);
    \draw[solid] (A) -- ++(\len,\len);
    \draw[solid] (A) -- ++(-\len,\len);
    \draw[solid] (A) -- ++(\len,-\len);
    \node[] at ($ (A) + (\eqdist,0) $) {$=-\llat_{a}$};

    \coordinate (A) at (2*\dist,0);
    \draw[dashed] (A) -- ++(-\len,-\len);
    \draw[dashed] (A) -- ++(\len,\len);
    \draw[dashed] (A) -- ++(-\len,\len);
    \draw[dashed] (A) -- ++(\len,-\len); 
    \node[] at ($ (A) + (\eqdist,0) $) {$=-\llat_{b}.$};
  \end{tikzpicture}
\end{figure}

The lattice propagators are given by
\begin{equation}
  \tilde{G}_{\alpha}(k) = \frac{1}{\hat{k}^{2} + \mlat^{2}_{\alpha}},
\end{equation}
where
\begin{equation}
  \klat^{2} = \sum_{i}\left( \frac{5}{2} - \frac{8}{3} \cos{(\klat_{i})} + \frac{1}{6}\cos{(2 \klat_i)} \right).
\end{equation}
We also use the following shorthand for the (dimensionless) lattice momentum integrals:
\begin{equation}
 \lint{BZ} := \int\limits_{[-\pi, \pi]^{2}} \frac{\mathrm{d}^2\klat}{(2\pi)^2}.
\end{equation}
For $\phi_{a}$ the only divergent diagrams are
\begin{figure}[htbp]
  \begin{tikzpicture}
    \begin{feynman}
      \diagram [small, horizontal=a to b, layered layout] {
        a--b--c
      };
      \path (b)--++(90:0.5) coordinate (A);
      \draw[dashed] (A) circle (0.5);

      \diagram [xshift=3cm,small, horizontal=a to b, layered layout] {
        a--b--c
      };

      \path (b)--++(90:0.5) coordinate (A);
      \draw[solid] (A) circle (0.5);
      \node[] at ($ (A) + (1.4,-0.5) $) {,};

    \end{feynman}
  \end{tikzpicture}
\end{figure}

\noindent both with the symmetry factor of $1/2$. Both are proportional to an integral of the form
\begin{equation}
  I_{\alpha}^{1l} \coloneqq \lint{BZ} \frac{1}{\klat^{2}+\mlat_{\alpha}^{2}}
\end{equation}
which is $\log$-divergent as $\mlat_{\alpha} \rightarrow 0$. The divergent parts can be extracted by the decomposition
\begin{align}
  \lint{BZ} \frac{1}{\klat^{2}+\mlat_{\alpha}^{2}} &= \lint{BZ} \left(\frac{1}{\klat^{2}+\mlat_{\alpha}^{2}} - \frac{1}{\kco^{2}+\mlat_{\alpha}^{2}}\right)  \label{eq:decomp-i} \\
                                            & + \lint{BZ} \frac{1}{\kco^{2}+\mlat_{\alpha}^{2}}, \label{eq:decomp-ii}
\end{align}
where $\kco^{2}$ is the continuum form $\klat_x^{2} + \klat_{y}^{2}$. The first term is finite for all $\mlat_{\alpha}$ and we can expand it as
\begin{align}
  &\lint{BZ} \left(\frac{1}{\klat^{2}+\mlat_{\alpha}^{2}} - \frac{1}{\kco^{2}+\mlat_{\alpha}^{2}}\right) \\
  =  &\lint{BZ} \left(\frac{1}{\klat^{2}} - \frac{1}{\kco^{2}}\right) + \mathcal O(\mlat_{\alpha}^{2}) \label{eq:decomp-i-expanded}
\end{align}
The divergences are now contained in the second integral \ref{eq:decomp-ii}. We decompose the domain into a sphere of radius $\pi$ surrounding the origin\footnote{Any radius $0 < \delta \leq \pi$ would do. A particular choice of $\delta$ makes the $\delta$-independence of the decomposition more evident}  and its complement in the Brillouin zone:
\begin{equation}
  \lint{BZ} \frac{1}{\kco^{2}+\mlat_{\alpha}^{2}} = \lint{|\kco| < \pi} \frac{1}{\kco^{2} + \mlat_{\alpha}^{2}} + \lint{|\kco| > \pi} \frac{1}{\kco^{2} + \mlat_{\alpha}^{2}}.
\end{equation}
We can compute the first integral directly using spherical coordinates, obtaining
\begin{align}
  &\frac{1}{4\pi}\left( \log\left(\pi^{2} + \mlat_{\alpha}^{2}\right) - \log\left(\mlat_{\alpha}^{2}\right) \right) \\
  = &\frac{1}{4\pi}\left(\log\left(\pi^{2}\right) - \log\left(\mlat_{\alpha}^{2}\right)\right) + \mathcal O(\mlat_{\alpha}^{2}). \label{eq:decomp-ii-expanded-a}
\end{align}
The second integral is well-behaved for all $\mlat_{\alpha}$ and we can expand in powers of $\mlat_{\alpha}$ as
\begin{equation}
  \lint{|\kco| > \pi} \frac{1}{\kco^{2} + \mlat_{\alpha}^{2}} = \lint{|\kco| > \pi} \frac{1}{\kco^{2}} + \mathcal O(\mlat_{\alpha}^{2}) \label{eq:decomp-ii-expanded-b}
\end{equation}
As we collect the terms \ref{eq:decomp-i-expanded}, \ref{eq:decomp-ii-expanded-a}, and \ref{eq:decomp-ii-expanded-b}, we obtain the result
\begin{align}
   I_{\alpha}^{1l} &= \lint{BZ} \left(\frac{1}{\klat^{2}} - \frac{1}{\kco^{2}}\right) + \lint{|\kco| > \pi} \frac{1}{\kco^{2}} + \frac{1}{2\pi}\log\left(\pi\right)\\
  &- \frac{1}{4\pi}\log\left(\mlat_{\alpha}^{2}\right) + \mathcal O(\mlat_{\alpha}^{2}).
\end{align}
The second row now contains the divergent behaviour at small $\mlat_{\alpha}$, while the first row is a pure number, $\Sigma$, depending only on the form of square of the lattice momentum, $\klat^2$. We can therefore write
\begin{equation}
  \label{eq:lattice-2-point-function}
  I_{\alpha}^{1l} = \Sigma - \frac{1}{4\pi}\log(\mlat_{\alpha}^{2}) + \mathcal O (\mlat_{\alpha}^{2}),
\end{equation}
where $\Sigma = 0.22862431$ by numerical integration.

Now recall that in the continuum theory the 1-loop contribution to the two-point function is proportional to the integral
\begin{equation}
  I^{1l}_{\alpha,\cont} = \int \frac{\mathrm d^{d}k}{(2\pi)^{d}} \frac{1}{k^2 + m_{\alpha}^2},
\end{equation}
which in $d = 2 - 2 \epsilon$ yields the familiar expansion
\begin{equation}
  I^{1l}_{\alpha,\cont} = \frac{1}{4\pi} \left(\frac{1}{\epsilon} + \log(4 \pi \mathrm e^{-\gamma_{\mathrm E}}) + \log\left(\frac{\bar \mu^2}{m_{\alpha}^2}\right) + \mathcal O(\epsilon)\right),
\end{equation}
where $\bar \mu$ is the renormalization scale. In the $\overline{\mathrm{MS}}$-scheme the one-loop result for the renormalized mass $m_{a, \mathrm R}^2$ is
\begin{equation}
m_{a, \mathrm R}^2(\bar \mu) = m_a^2 - \frac{1}{8\pi} \left (\lambda_{a}^{\cont} \log\left(\frac{\bar \mu^2}{m_{a}^2}\right) + \mu^{\cont} \log\left(\frac{\bar \mu^2}{m_{b}^2}\right) \right).
\end{equation}
This also corresponds to the physical pole mass. On the lattice the corresponding quantity (in lattice units) is
\begin{equation}
\mlat_{a, \mathrm{phys}}^2 = \mlat_a^2 - \frac{1}{2}(\llat_{a} I_{a}^{1l} + \mulat I_{b}^{1l})  - \delta \mlat^2_{a}.
\end{equation}
We now require that these two match. Recalling that the difference between the lattice and continuum quartic couplings is of $\mathcal O(a^2)$, the choice
\begin{equation}
\delta \mlat^2_{a} = - \frac{a^2}{2}(\lambda^{\cont}_{a} + \mu^{\cont}) \left( \Sigma - \frac{1}{4\pi} \log \left( a^2 \bar \mu^2 \right)\right)
\end{equation}
is sufficient to ensure that the lattice and continuum theories match in the continuum limit up to $\mathcal O (a^2)$ errors.

For $\phi_{b}$ the process is of course identical.

% --------------------
% Bibliography
% --------------------
\bibliographystyle{apsrev4-2}
\bibliography{biblio-arxiv}

\end{document}